\documentclass[pra,amsmath,superscriptaddress,twocolumn,showpacs]{revtex4-1}
\usepackage{bm}
\usepackage{amssymb}
\usepackage{colordvi}
\usepackage{graphicx}
\usepackage{color}
\usepackage{hyperref}

\newcommand{\be}{\begin{equation}}
\newcommand{\ee}{\end{equation}}
\newcommand{\bea}{\begin{eqnarray}}
\newcommand{\eea}{\end{eqnarray}}

\newcommand{\vep}{\varepsilon}
\newcommand{\ave}[1]{\langle #1\rangle}
\newcommand{\ome}{\omega}

\def\nn{\nonumber}

\begin{document}

\title{Enhancing thermoelectric performance using nonlinear transport effects}

\author{Jian-Hua Jiang}
\affiliation{College of Physics, Optoelectronics and Energy, \&
  Collaborative Innovation Center of Suzhou Nano Science and
  Technology, Soochow University, 1 Shizi Street, Suzhou 215006,
  China}
\author{Yoseph Imry}
\affiliation{Department of Condensed Matter Physics, Weizmann Institute of
  Science, Rehovot 76100, Israel}

\date{\today}

\begin{abstract}
  We study nonlinear transport effects on the maximum efficiency and
  power for both inelastic and elastic thermoelectric generators. The
  former refers to phonon-assisted hopping in double quantum-dots, while
  the latter is represented by elastic tunneling through a single 
  quantum-dot. {We find that nonlinear thermoelectric
    transport can lead to enhanced efficiency and power for both types
    of devices. A comprehensive survey of various
    quantum-dot energy, temperature, and parasitic heat conduction reveals 
    that the nonlinear transport induced 
    improvements of the maximum efficiency and power are overall much more
    significant for inelastic devices than the elastic devices, even for
    temperature biases as small as $T_h=1.2T_c$ ($T_h$ and $T_c$ are the
    temperatures of the hot and cold reservoirs, respectively). The
    underlying mechanism is revealed as due to the fact that, unlike the
    Fermi distribution, the Bose distribution is not bounded when
    the temperature bias increases.} A large flux density of absorbed 
  phonons leads to great enhancement of the electrical current, the
  output power, and the energy efficiency, dominating over the
  concurrent increase of the parasitic heat current.
  {Our study reveals that nonlinear transport effects
    can be a useful tool for improving thermoelectric performance.}
\end{abstract}

\pacs{05.70.Ln, 84.60.-h, 88.05.De, 88.05.Bc}

\maketitle

\section{Introduction}
Thermoelectric energy conversion is important both for fundamental
research on, e.g., the physics of mesoscopic electron
systems\cite{joe} and for applications\cite{book}. Advancements in
material sciences and nanotechnology have been pushing the frontiers
of thermoelectric researches. Nanostructured bulk thermoelectric
materials\cite{hicks,nano1,nano2} have yielded a (device) figure of
merit as high as $ZT\simeq 1.34$\cite{kanatzidis,ncomm}, reaching to
$\simeq$28\% of the Carnot efficiency. In the past few years, studies
on inelastic thermoelectricity 
opened a unprecedented avenue for potential high-performance
thermoelectrics\cite{joe1,sanchez1,jiang1,o0,DS,sanchez2,magnon,jiang2,pn,o1,sanchez3,jordan,o2,3tjap,jiang4,nonlinear,jiang5,o3,felix,o4,exp1,exp2,exp3,lijie,rev}.
For an inelastic thermoelectric generator, the input heat is supplied
in the form of phonons or other bosons that assist the inelastic 
transport\cite{jiang1,rev}. As a consequence, the figure of merit can be
written as\cite{pn,rev} 
\be
ZT = \frac{\ave{\ome}^2}{\ave{\ome^2}-\ave{\ome}^2+\Delta}, \label{fom}
\ee
where $\ome$ is the energy of the bosons (throughout this paper we set
$\hbar\equiv 1$). Here the average is weighted by
the contribution of each inelastic transport channel to the 
electrical conductance\cite{pn,rev}. The term $\Delta$ describes the
reduction of the figure of merit due to the parasitic heat
conduction\cite{pn,rev}. Note that elastic transport processes (i.e.,
$\ome=0$) do not contribute to the inelastic
thermoelectricity\cite{jiang1}, although they lead to 
the conventional thermoelectricity.

Inelastic thermoelectricity has several advantages over the
conventional thermoelectricity that is based on elastic transport.
First, for inelastic thermoelectrics, a large figure of
merit can be achieved by a small bandwidth of the bosons that assist
the inelastic transport\cite{pn},
instead of a small bandwidth of electrons as in Mahan and Sofo's
proposal for the conventional thermoelectricity\cite{ms}. The
small bandwidth of the bosons that are involved in the inelastic
transport processes does 
{\em not} suppress the electrical conductance (unlike in Mahan and
Sofo's proposal\cite{zhou}). Significant inelastic transport can be
realized in ionic semiconductors where the electron-phonon scattering
is very strong\cite{koch,gan2}. Second, in conventional
thermoelectricity, heat and charge are transported in the same
direction which makes it rather difficult to improve the figure of
merit by reducing the parasitic phonon thermal conductivity while
maintaining high electrical
conductivity\cite{zhou,book,nano1,nano2,kanatzidis}. If heat and
charge are transported separately, it will be much easier to
manipulate the parasitic phonon thermal conductivity and electrical
conductivity concurrently in the aim of high-performance
thermoelectrics\cite{pn,3tjap}. It has been shown that inelastic
thermoelectric devices based on $p$-$n$ junctions with very small band  
gap (e.g., HgCdTe $p$-$n$ junctions with band gap $\simeq
50$~meV)\cite{pn} or quantum-dots (QDs) arrays\cite{3tjap} can have
rather high figure of merit even when the phonon parasitic heat
conductivity is taken into consideration. It was also shown that in
these devices, the inelastic thermoelectricity has larger figure of
merit and power factor than the conventional thermoelectricity\cite{3tjap}.

In this work, we show that nonlinear transport effects can further
enhance inelastic thermoelectric efficiency and power when the voltage
and/or temperature bias is large. Specifically, when the temperature
of the phonon bath is increased, the nonlinear thermoelectric
transport leads to significant improvement of both the
heat-to-work energy efficiency and the output
electric power. All these effects are found to be
associated with inelastic thermoelectric transport. In contrast, for
elastic thermoelectric transport the nonlinear effects lead to
marginal improvement or reduction of the maximum efficiency. The
effects of parasitic heat conduction in both the inelastic and 
elastic thermoelectric devices are investigated carefully. We find
that the key difference between the elastic (or the ``conventional'') and
the inelastic thermoelectricity is that at large temperature biases the
inelastic thermoelectric currents increase dramatically due to the
exponentially large Bose distribution factor. In contrast, the elastic
transport currents are limited by the Pauli exclusion
principles. In physical terms, inelastic transport currents are
proportional to the flux of bosons absorbed by electrons and enhanced
by the Bose distribution, whereas the elastic transport currents
around the chemical potential are limited by the Fermi distributions.
Considering the similarity between the inelastic
thermoelectrics\cite{pn,rev} and solar
cells\cite{shockley,solar,booksc}, our theory may also be regarded as
a possible answer to the question why solar cells have much higher
record efficiency than thermoelectrics (even with the Carnot
efficiency $\eta_C$ taking into account, i.e., considering the ratio
$\eta/\eta_C$)\cite{book,shockley,solar,booksc}. 

Although nonlinear effects in
thermoelectric transport have been discussed in previous studies for
both elastic\cite{n1,n2,n3} and inelastic transport\cite{n1-3t,n2-3t},
it is our focus here to study the effects of the nonlinear transport on
the maximum energy efficiency and the maximum output power.

\section{Double Quantum-Dots Three-Terminal Thermoelectric Transport}

\begin{figure}[htb]
  \includegraphics[width=8.4cm]{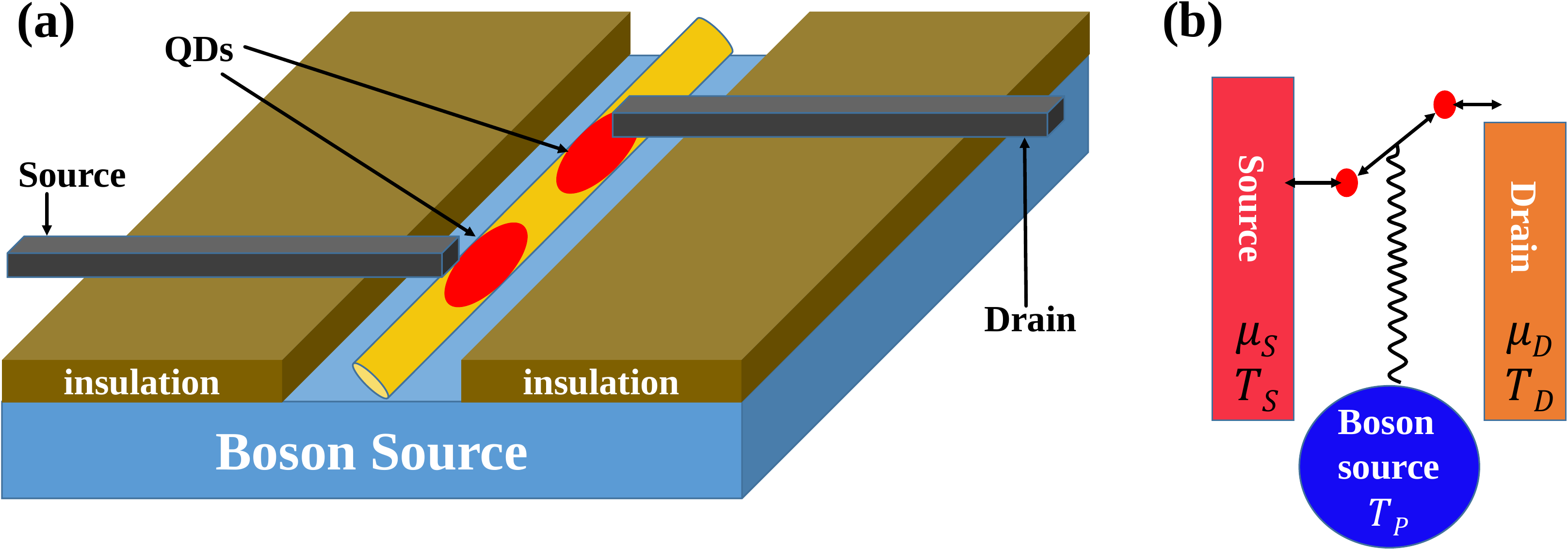}
  \caption{(Color online) (a) Schematic of a double QDs three-terminal
    thermoelectric device. Two QDs are embedded in a nanowire which is
    placed on a substrate. The substrate acts as a boson source that
    emits bosons (e.g., phonons) to assist inelastic hopping between
    the two QDs. The source electrode (isolated from
    the substrate by thermally insulating layers) is in contact with
    one QD, while the drain electrode is in contact with the
    other QD. The QD connected to the drain has a higher energy than
    the QD connected to the source. (b) Illustration of the
    three-terminal inelastic transport. An electron left the source
    into the first QD (with energy $E_1$) hops to the second QD (with
    a different energy $E_2$) as assisted by a boson (e.g., a phonon)
    from the boson source (with temperature $T_P$). The electron then
    tunnels into the drain electrode from the second QD. Such a
    process gives inelastic charge transfer from the source to the
    drain assisted by the boson from the boson source. Both the
    process and its time-reversal contribute to the inelastic
    thermoelectricity in the system. The electrochemical potential and
    temperature of the source (drain) are $\mu_S$ and $T_S$ ($\mu_D$
    and $T_D$), respectively.}
\end{figure}

We study the energy efficiency and power for a double-QDs
three-terminal thermoelectric device in the nonlinear transport 
regime. The device is schematically depicted in Fig.~1(a) and
explained in the caption. Two QDs, with energy $E_1$ and $E_2$, are
embedded in a nanowire. {For simplicity, we consider
  the situations where there is only one spin-degenerate energy level
  in each QD relevant for thermoelectric transport. This assumption
  can be justified for small QDs where the energy level separation in
  each QD is much larger than  $k_BT$\cite{ganqd}. The
  difference between $E_1$ and $E_2$ can be tuned by the local
  potentials of the two QDs.} We
consider high-temperature transport where the effect of Coulomb
interaction is negligible. The {electron} spin 
degeneracy doubles the electronic charge and heat currents. The phonon
assisted hopping transport is illustrated in Fig.~1(b) and explained
in the caption. Besides, electron can also tunnel elastically between
the source and the drain via the two QDs. The sequential
tunneling processes are suppressed when the energy difference
$|E_2-E_1|$ is significant. A similar system was studied in
Ref.~\cite{nonlinear} where thermoelectric rectification and
transistor effects were obtained. The double-QDs device is described
by the following Hamiltonian 
\be
 H = H_{QD} + H_{ep} + H_p + H_{tun} + H_{lead} , 
\ee
where the QDs are described by 
\be
 H_{QD} = E_1 d_1^\dagger d_1 + E_2 d_2^\dagger d_2 + t (d_1^\dagger
d_2 + d_2^\dagger d_1) ,
\ee
the inter-QD electron-phonon interaction is given by
\be
 H_{ep} = \sum_{\vec{q}} M_{\vec{q}} d_1^\dagger d_2 (a_{\vec{q}}+a_{-\vec{q}}^\dagger) + {\rm H.c.}
,
\ee
with the phonon Hamiltonian
\be
H_{p} = \sum_{\vec{q}} \ome_{\vec{q}} (a_{\vec{q}}^\dagger
a_{\vec{q}} + \frac{1}{2}).
\ee
The tunneling between the QDs and the electrodes are described by
\be
H_{tun} = \sum_{\vec{k}} (V_{S,\vec{k}} c_{S,\vec{k}}d_1^\dagger + V_{D,\vec{k}}
c_{D,\vec{k}}d_2^\dagger) + {\rm H.c.} ,
\ee
where the two leads (the source and the drain) have the Hamiltonian,
\be
H_{lead} = \sum_{j=S,D}\sum_{\vec{k}} \vep_{j,\vec{k}}
c_{j,\vec{k}}^\dagger c_{j,\vec{k}} .
\ee
Here $d_i^\dagger$ ($i=1,2$) creates an electron in the $i^{th}$ QD,
$c_{S,\vec{k}}^\dagger$ and $c_{D,\vec{k}}^\dagger$ create electrons in the
source and the drain with wavevector $\vec{k}$,
respectively. $a_{\vec{q}}^\dagger$ creates a phonon with wavevector
$\vec{q}$ in the nanowire. For simplicity, the spin index of electrons and
the branch index of phonons are omitted in the above equations. $t$ is the
tunnel coupling between the two QDs. $M_{\vec{q}}$ is the electron-phonon
coupling matrix element. The phonons here should be phonons in the
nanowire. We assume that the thermal contact between the nanowire and
the substrate is good so that their temperatures are approximately the
same (denoted as $T_P$). The frequency of the phonon is 
$\ome_{\vec{q}}$. The tunnel coupling between the first (second) QD
and the source (drain) is described by $V_{S,\vec{k}}$
($V_{D,\vec{k}}$). Both the coupling and the spectrum in the leads
$\vep_{j, {\vec k}}$ ($j=S,D$) determine the tunneling rate between
the electrodes and the QDs. We shall describe the tunneling between
the source (drain) and the first 
(second) QD phenomenologically as an 
energy-independent constant, $\gamma_1$ ($\gamma_2$). For simplicity,
we consider the situations with $\gamma_1=\gamma_2=\gamma_e$. The
tunneling between the first (second) QD and the drain (source) can be
obtained by the perturbation theory as [for $t^2\ll (E_2-E_1)^2$]
\begin{subequations}
\begin{align}
&\gamma_1^\prime=\gamma_2^\prime\equiv \gamma_e^\prime =
\frac{t^2\gamma_{e}}{ (E_2-E_1)^2 } ,
\end{align} 
\end{subequations}
The rate of electron transfer from the first QD to the second QD due
to the electron-phonon scattering is given by
\be
\Gamma_{12} = \gamma_{ep} [f_1 (1-f_2) N_P^+ - f_2(1-f_1) N_P^-]
\ee
where
$\gamma_{ep}=2\pi\sum_{\vec{q}}|M_{\vec{q}}|^2\delta(|E_1-E_2|-\ome_{\vec{q}})$
is the electron-phonon scattering rate calculated from the Fermi
golden rule. $f_i$ ($i=1,2$) are the probabilities of finding an
electron on the $i$-th QD. $N_P^\pm={\rm Abs}(n_B[\pm (E_2-E_1)/(k_BT_P)])$
where $n_B(x)=1/(e^x-1)$ is the Bose distribution function.

This model has been studied before by the
authors\cite{jiang1,jiang2,nonlinear,rev}. Particularly, 
in Ref.~\cite{nonlinear}, we studied the nonlinear thermoelectric
transport in the system and shown that the device can function as
cross-correlated thermoelectric rectifications and transistors. One
of the remarkable properties of the system is that it allows thermal
transistor effect without relying on negative differential thermal
conductance, in contrast to common believes. Here we focus on the
thermoelectric efficiency and power of the device in the nonlinear
regime. {While Ref.~\cite{nonlinear} deals with the effects of
low-frequency phonons at low temperatures, here we focus on the
situations where the phonon energy $\ome$ is close to $k_BT_D$ ($T_D$
is the Debye temperature) for elevated temperatures. There are at
least two advantages in this regime: First, the electron-phonon
interaction is much stronger, leading to higher current
density in the inelastic transport channels; Second, the phonon
energy, $\sim k_BT_D$, is usually higher than $k_BT$, which is
beneficial for the figure of merit, as indicated by Eq.~(\ref{fom}).}

There are two electronic and one bosonic reservoirs in our
thermoelectric device: the source (with chemical potential $\mu_S$ and
temperature $T_S$), the drain (with chemical potential $\mu_D$ and
temperature $T_D$), and the phonon bath (i.e., the substrate, with
temperature $T_P$). In this three-terminal device, all reservoirs are
isolated using thermal insulation [see Fig.~1(a)]. Energy and charge
are transported through the double QDs. Thermodynamic analysis gives
the following currents and their conjugated affinities,\cite{nonlinear}
\begin{align}
& I_e = - e \frac{d N_S}{dt}, \quad A_e =
\frac{\mu_S-\mu_D}{e}(\frac{1}{2T_S}+\frac{1}{2T_D}) ,\nn \\
& I_{Q,e} = \frac{1}{2} (\frac{d Q_D}{dt} - \frac{d Q_S}{dt} ) , \quad
A_{Q,e} = \frac{1}{T_D} -\frac{1}{T_S} ,\nn\\
& Q_{in} = - \frac{d Q_{P}}{dt} ,\quad A_{in} =
\frac{1}{2T_S}+\frac{1}{2T_D} - \frac{1}{T_P} .
\end{align}
Here $e<0$ is the electron charge and $N_S$ is the electron number in
the source. $Q_S$, $Q_D$, and $Q_P$ represent the heat content
[$Q_i=E_i-\mu_i N_i$ for $i=S,D$, and $Q_P=E_P$ where $E_i$
($i=S,D,P$) are the total energy and $N_i$ ($i=S,D$) are the total
electron number for these reservoirs] for the source,
drain, and phonon bath, respectively. In this work, we focus on the
situation with $T_S=T_D\equiv T_e$. Thus we have
\begin{subequations}
\begin{align}
& A_e = \frac{V}{T_e},\quad V\equiv \frac{\mu_S-\mu_D}{e} ,\\
& A_{Q,e}=0, \quad A_{in} = \frac{1}{T_e} - \frac{1}{T_P} .
\end{align}
\end{subequations}
The chemical potentials of the source and the drain are set
anti-symmetrically around the equilibrium value $\mu\equiv 0$, i.e.,
$\mu_S=eV/2=-\mu_D$.

We consider harvesting the heat from the (hot) phonon bath to generate
electricity. The energy efficiency is hence
\be
\eta = \frac{-I_eV}{Q_{in}} \le \eta_C \equiv \frac{T_h - T_c}{T_h}
, 
\ee
where $\eta_C$ is the Carnot efficiency, $T_h=T_P$ and $T_c=T_e$. Both
the efficiency $\eta$ and the output power 
\be
P=-I_eV
\ee
are of central
concern in this work. The heat injected into the system from the boson
source (here is the phonon bath\cite{jiang1}) is 
\be
Q_{in} = I_{Q,ph} + I_{Q,pr} , \quad I_{Q,ph} \equiv 2\ome\Gamma_{12}, \quad
\ome\equiv E_2-E_1 \nn
\ee
where $I_{Q,pr}$ is the parasitic phonon heat current. The factor of two
in the above equation comes from electron spin degeneracy.

The steady-state transport in the three-terminal device is studied via
the following rate equations which treat the elastic and inelastic
processes on equal footing,
\begin{subequations}
\begin{align}
& 0 = \frac{d f_1}{dt} = \nn \\ 
&\quad - \gamma_e [ f_1 - f_S(E_1)] -  \gamma_e^\prime [ f_1
- f_D(E_1)] - \Gamma_{12} , \\
& 0 = \frac{d f_2}{dt} = \nn \\
&\quad - \gamma_e [ f_2 - f_D(E_2) ] -
\gamma_e^\prime [ f_2 - f_S(E_2) ] + \Gamma_{12} .
\end{align}
\end{subequations}
By solving the above equations, we can determine the nonequilibrium
steady-state distributions on the QDs, $f_1$ and $f_2$.
Afterwards, the electrical and heat currents can be obtained via the
following equations (the factor of two originates from electron-spin
degeneracy) 
\begin{subequations}
\begin{align}
& I_e = 2 e \gamma_e [ f_S(E_1) - f_1] + 2 e \gamma_e^\prime [
f_S(E_2) - f_2] ,\\
& Q_{in} = 2\ome\Gamma_{12} + I_{Q,pr} ,
\end{align}
\label{fullC}
\end{subequations}
where the parasitic phonon heat current, $I_{Q,pr}$, is calculated via 
\be
I_{Q,pr} = \int_0^\infty \frac{dE}{2\pi} E {\cal T}_{pr}(E) [n_B(\frac{E}{k_BT_P}) -
n_B(\frac{E}{k_BT_e}) ] . \label{pr}
\ee
Here ${\cal T}_{pr}(E)$ is the energy-dependent transmission
function for phonons. We note that serious
considerations of the parasitic heat conduction in the literature are
scarce, particularly in the nonlinear regime (Note that a
phenomenological heat conductivity parameter cannot include the
nonlinear effect). This is mainly because
such an effect depends on the details of the device and contacts. Hence
it is rather difficult to deduce the transmission function from
theoretical aspects (if it is at all possible). To avoid the complexity, we
include the parasitic heat conduction via an ideally simple 
transmission function for phonons transferring between the phonon
bath and the source/drain terminals, ${\cal T}_{pr}(E)=\alpha
\Theta(E_{cut}-E)$ where $\alpha$ is a dimensionless constant and
$E_{cut}$ is the cut-off energy of the phonons. These phonons are
{\em not} absorbed by electrons and hence do not contribute to
the thermoelectric energy conversion. The parameters
$\alpha$ and $E_{cut}$ measure the average transmission level
and the effective bandwidth of the parasitic phonons, respectively.
The parasitic heat current
measures how much energy flux carried by the bosons that are {\em not}
absorbed by the electrons (i.e., not involved in the inelastic transport).

The linear transport coefficients are obtained by calculating the
ratios between currents and affinities in the regime with very small
voltage bias and temperature difference. In the linear-response regime
the current-force relations are
\be
I_i =\sum_j M_{ij} A_j \label{linear}
\ee
for the charge and heat transports. In the following
we will compare the energy efficiency, power, and currents between
the full calculation using Eqs.~(\ref{fullC}) (denoted as ``nonlinear''
for short)
and the calculation using Eq.~(\ref{linear}) and assuming its validity
for large biases (denoted as ``linear'' for short).


\section{Nonlinear transport enhances efficiency and power for
  inelastic thermoelectric devices}

\begin{figure}[htb]
  \includegraphics[width=4.35cm]{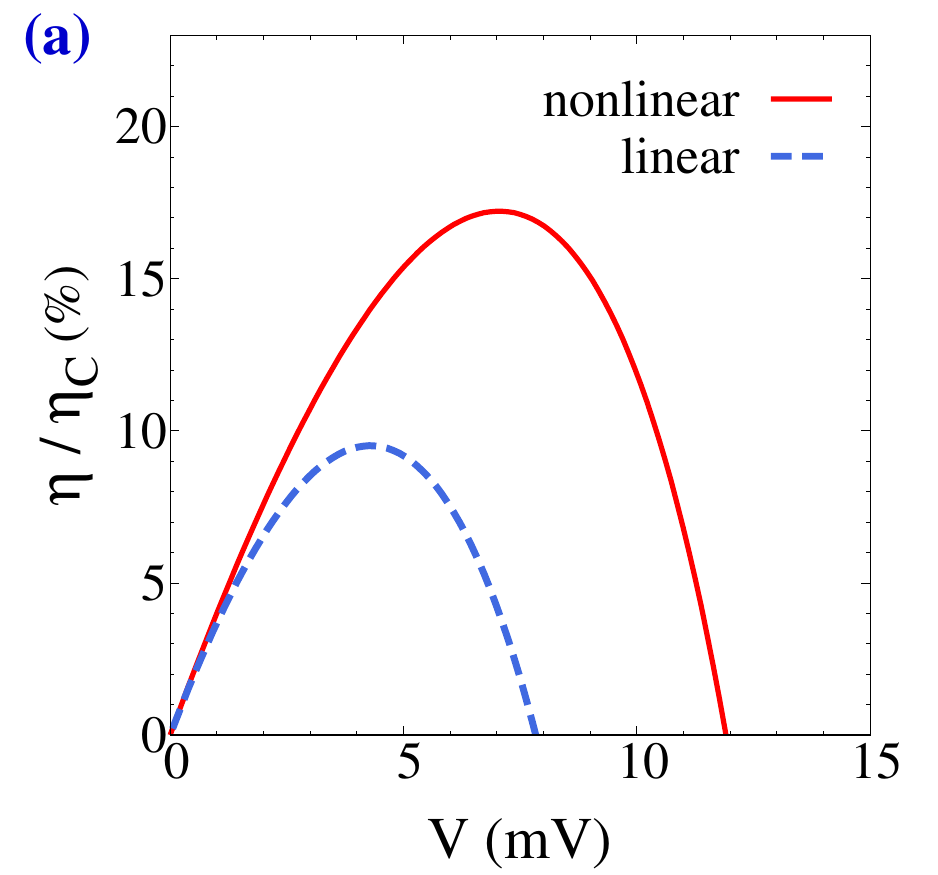}\includegraphics[width=4.35cm]{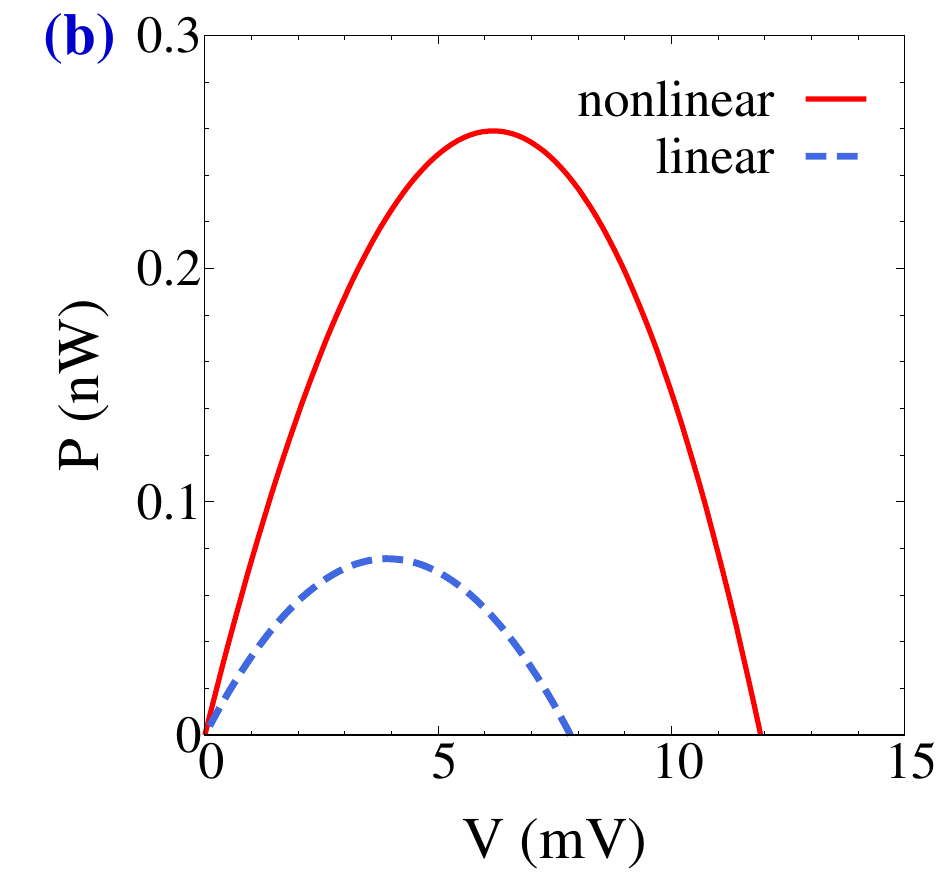}
  \includegraphics[width=4.35cm]{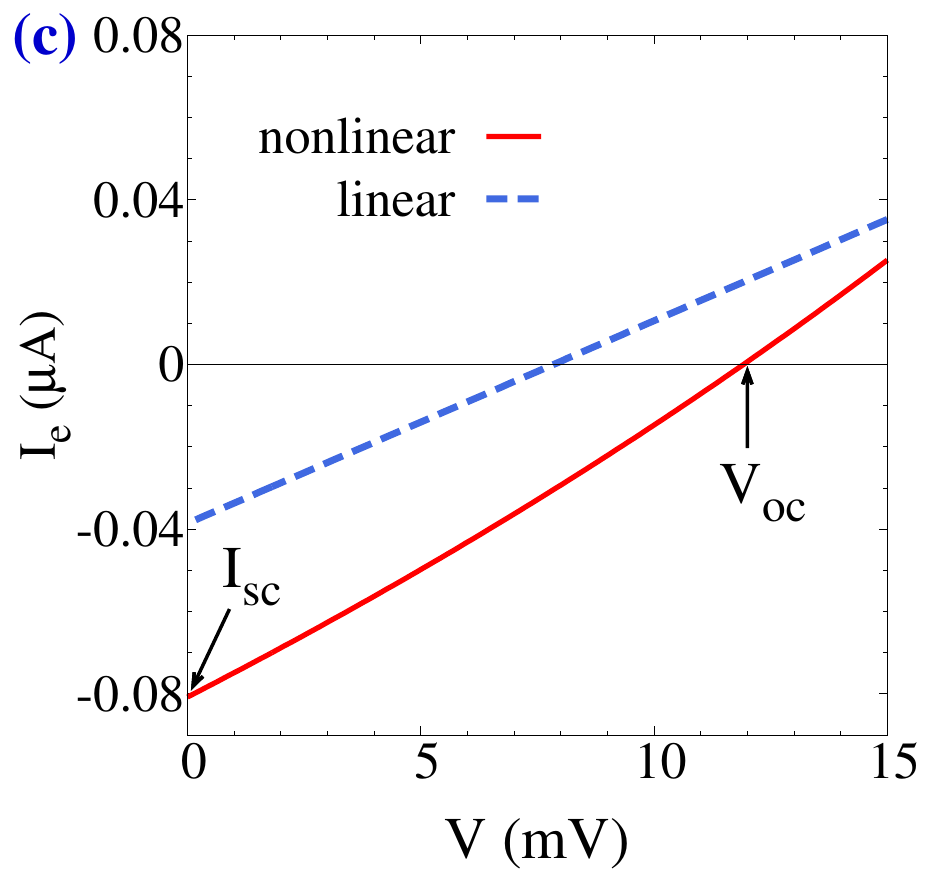}\includegraphics[width=4.35cm]{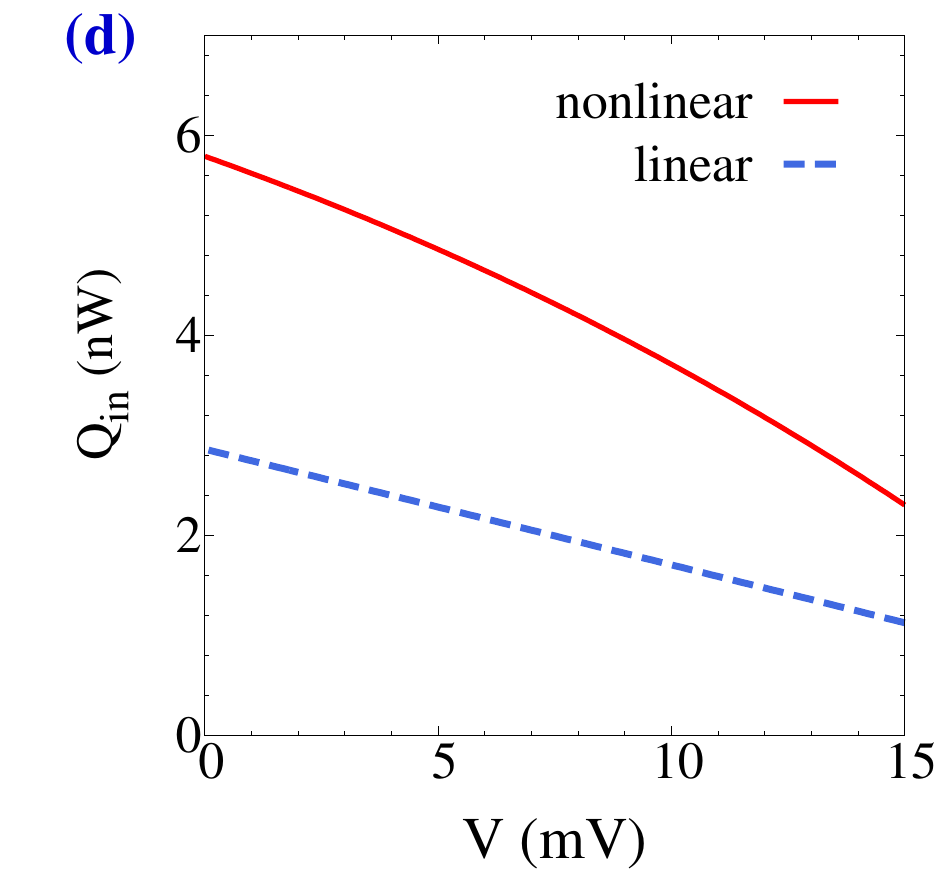}
  \includegraphics[width=4.35cm]{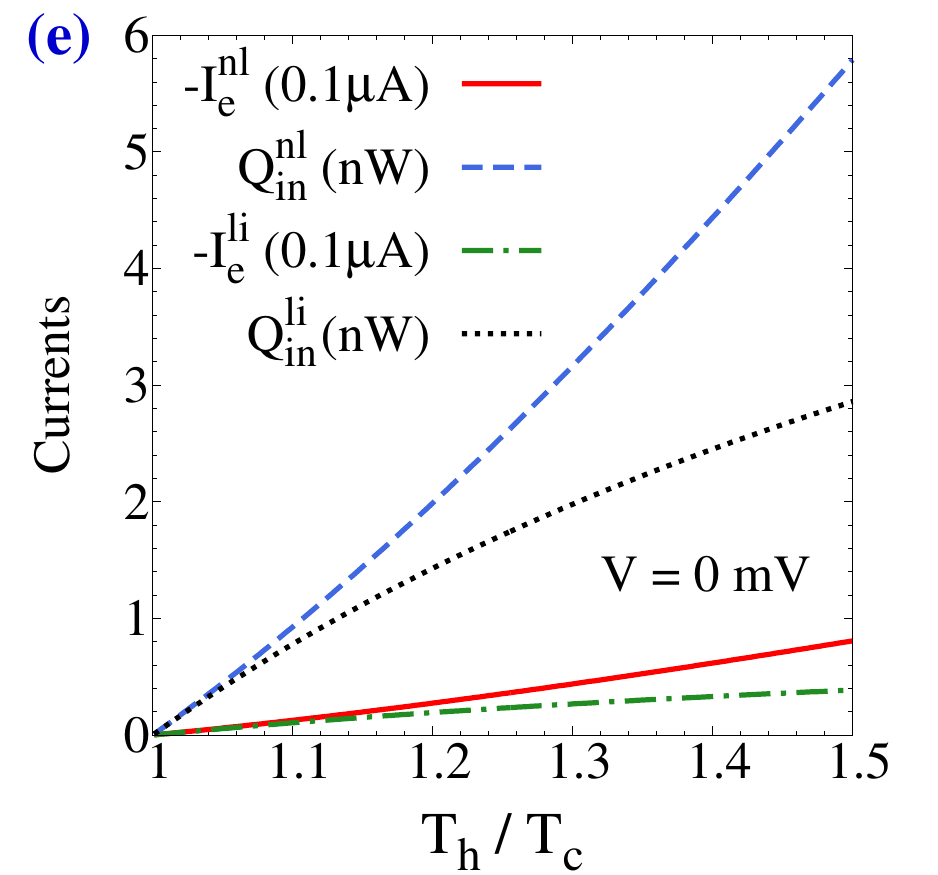}\includegraphics[width=4.35cm]{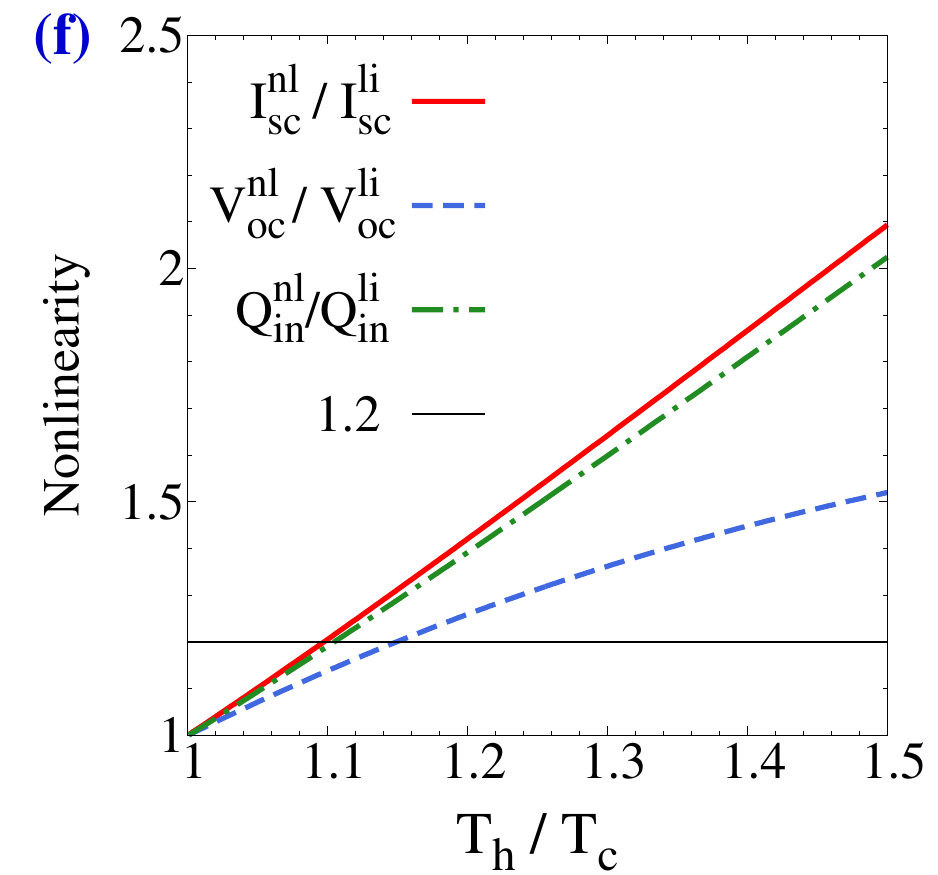}
  \caption{(Color online) Inelastic thermoelectricity. (a) Energy efficiency $\eta$ in units of the
    Carnot efficiency $\eta_C$ and (b) output power $P$ as
    functions of voltage $V$ (in units of mV). Parameters:
    $\gamma_e=10$~meV, $\gamma_{ep}=5$~meV, $t=10$~meV,
    $E_{cut}=100$~meV, $T_h=225$~K, $T_c=150$~K, $\alpha=0.1$,
    $E_1=-E_2=-2k_BT_c$. {(c) Electrical and (d) heat currents obtained
    via full calculation using Eq.~(\ref{fullC}) (labeled as
    ``nonlinear'') and those calculated via the linear-approximation
    using Eq.(\ref{linear}) (labeled as ``linear''). (e) The
    electrical currents $I_e$ and heat currents $Q_{in}$ as functions
    of $T_h/T_c$ for the short-circuit condition $V=0$. Here the
    superscript ``nl'' denotes the results from full calculation with
    nonlinear effects, while the superscript ``li'' denotes the
    results from linear-approximation. The same notation of
    superscript is used in (f) as well. (f) The short-circuit current 
    $I_{sc}$, the open-circuit voltage $V_{oc}$, and the heat current
    $Q_{in}$ measured by their values in the linear-approximation for
    various $T_h/T_c$. The line ``1.2'' is used to signify
    considerable nonlinearity.}}
\end{figure}

{
We calculate the efficiency and output power for a three-terminal
inelastic thermoelectric generator. At fixed temperatures $T_h$ and
$T_c$ (with $T_h=225$~K and $T_c=150$~K), the nonlinear transport
yields significant improvement of the maximum efficiency and power, as
shown in Figs.~2(a) and 2(b). The maximum efficiency calculated via
linear-approximation is only $9.7\%$ of the Carnot efficiency, while
the full calculation (including the nonlinear transport effect) leads
to a maximum efficiency of $17$\% of the Carnot efficiency. The
maximum output power increases from 0.08~nW to 0.26~nW, when the
nonlinear transport effect is taken into account. The temperatures
adopted here is consistent with the experimental fact that large
temperature difference is easier to achieve at low
temperature\cite{glushk}. Nevertheless, our theory also works for high 
temperatures and give the same conclusions on the nonlinear
thermoelectric performance.}

{To understand the physical origin of the enhancement of the maximum
efficiency and power, we first study how the electrical and heat
currents are affected by the nonlinear transport effect. From Fig.~2(c) it is
seen that the electrical current is considerably enhanced due to
the nonlinear effect. Here we adopt two quantities used in the
solar cell literature\cite{booksc} to analyze the nonlinear
thermoelectric transport: the short-circuit current $I_{sc}$ (i.e.,
the electrical current at $V=0$) and the open-circuit voltage $V_{oc}$
(i.e., the voltage at which the electrical current vanishes). The
product of the two
\be
{\cal P}\equiv - I_{sc}V_{oc}
\ee
characterizes the
maximum output power\cite{booksc}. Indeed, the full calculation gives
a ${\cal P}$ more than 3 times as large as the ${\cal P}$ obtained
from the linear-approximation, agreeing well with the improvement of the
maximum power.}

{To understand the improvement of the maximum
  efficiency, we also need to examine how the input heat $Q_{in}$ is
  affected by the nonlinear transport effect. Fig.~2(d) shows that the
  input heat at $V=0$, $Q_{in}(V=0)$, is increased to about 2 times as
  large as that obtained via the linear-approximation. The increase of
  the output power then exceeds that of the input heat, explaining the
  improvement of the maximum efficiency. The above analysis reveals
  the importance of the quantities $I_{sc}$, $V_{oc}$, and
  $Q_{in}(V=0)$ in the study of the maximum power and efficiency,
  which will also be exploited in the study of the nonlinear 
  transport effect on the performance
  of the elastic tunneling thermoelectric device later.}

{The nonlinear transport effect is reflected directly
  in the dependences of the electrical and heat currents on the
  temperature ratio $T_h/T_c$ when $T_c$ is fixed, as shown in
  Fig.~2(e). Both the electrical and heat currents are much enhanced
  due to the nonlinear transport effect, even for $T_h\le 1.5T_c$. To
  manifest the nonlinear effect more obviously, we shown in Fig.~2(f)
  the ratio of the short-circuit current obtained via the full
  calculation (i.e., with the nonlinear transport effect) 
  $I_{sc}^{nl}$ to that obtained via the linear-approximation,
  $I_{sc}^{li}$, as well as similar ratios for the other two quantities,
  the open-circuit voltage $V_{oc}$ and the short-circuit heat current
  $Q_{in}$. It is seen that these ratios increase rapidly with
  increasing temperature $T_h$ when $T_c$ is fixed. For $T_h\ge 1.1
  T_c$, the nonlinear transport effect is already prominent enough to
  produce $>20$\% deviation between the ``nonlinear'' and the
  ``linear'' currents and voltage. Such enhancement of currents and
  voltage due to the nonlinear effect is the origin of the significant
  improvement of the performance of the inelastic thermoelectric generator.}

\section{Effects of nonlinear transport on efficiency and power for elastic
  thermoelectric devices}

\begin{figure}[htb]
  \includegraphics[width=4.35cm]{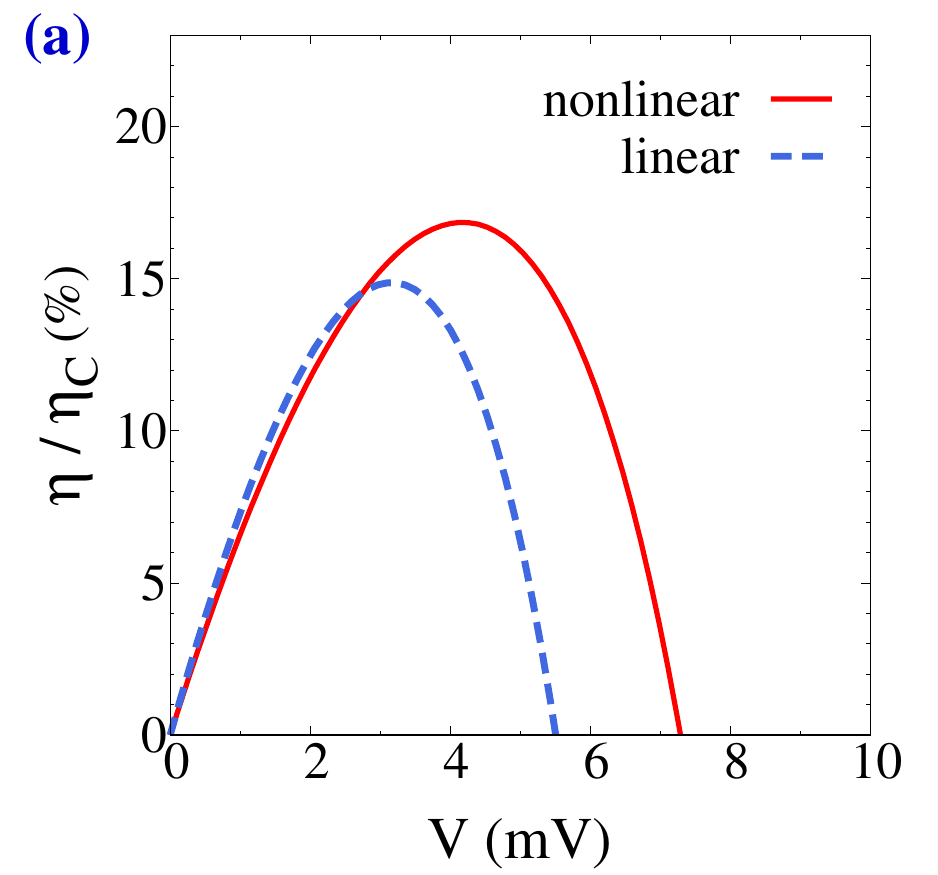}\includegraphics[width=4.35cm]{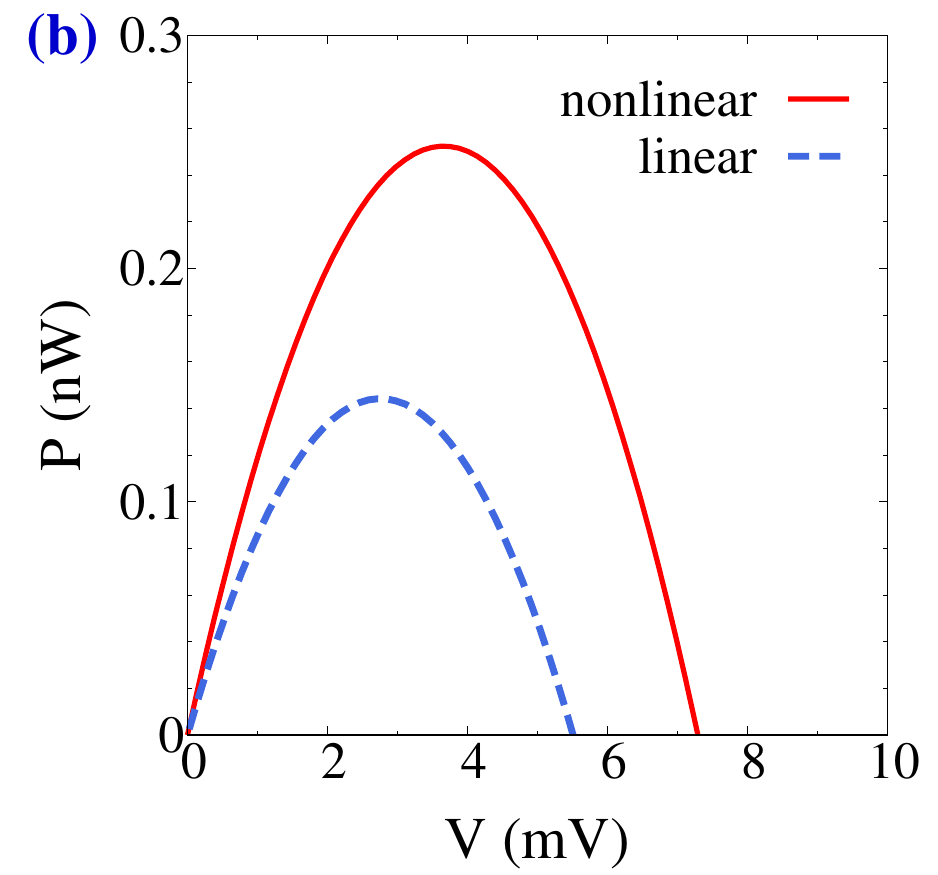}
  \includegraphics[width=4.35cm]{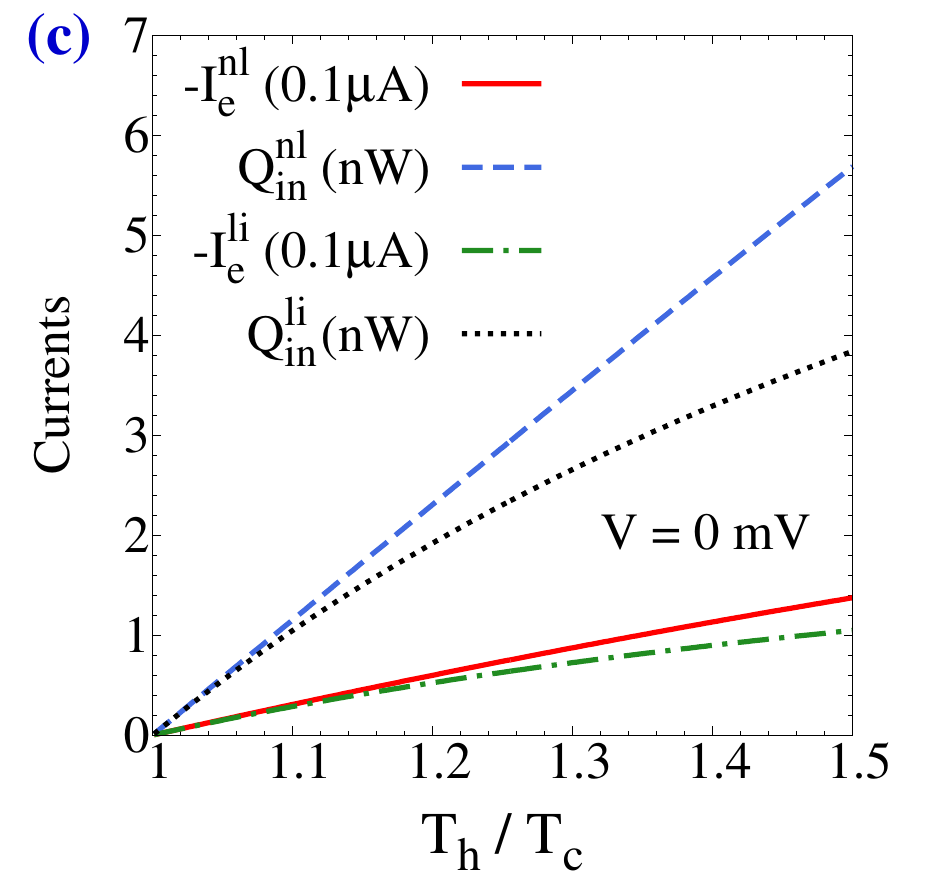}\includegraphics[width=4.35cm]{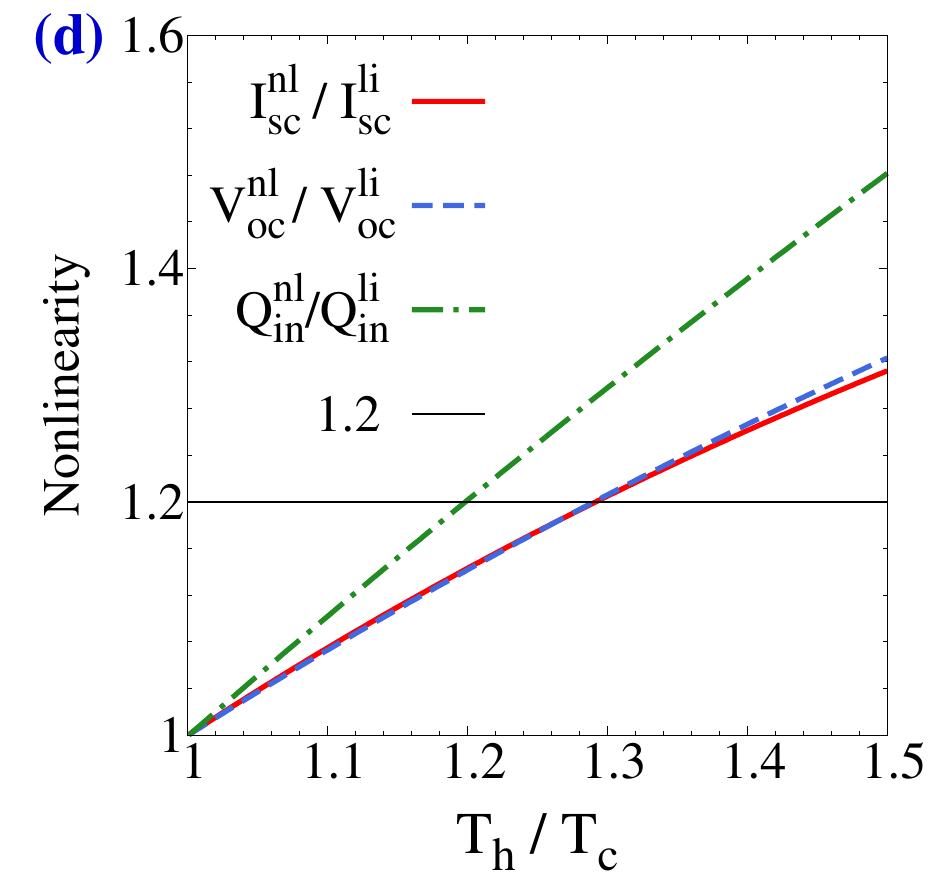}
  \caption{(Color online) Elastic thermoelectricity. (a) Energy
    efficiency $\eta$ in units of the Carnot efficiency $\eta_C$ and
    (b) output power $P$ as functions of voltage $V$ (in units of mV)
    for the elastic thermoelectric device. {Parameters:
      $\gamma_e=10$~meV, $E_{cut}=100$~meV, $T_h=225$~K, $T_c=150$~K,
      $\alpha=0.1$, $E_0=2k_BT_c$. (c) Electrical and heat currents as
      functions of the  temperature ratio $T_h/T_c$ for the
      short-circuit condition $V=0$. (d) The short-circuit current 
      $I_{sc}$, the open-circuit voltage $V_{oc}$, and the heat current
      $Q_{in}$ measured by their values in the linear-approximation
      vs. $T_h/T_c$. The line ``1.2'' is used to signify
      considerable nonlinearity. The quantities from the full calculation (i.e.,
      with nonlinear effects) are labeled by the superscript ``nl'',
      while those from the linear-approximation are denoted by the
      superscript ``li''.}}
\end{figure}

As a comparison, we also study the nonlinear transport effects on
the performance of elastic thermoelectric devices. A simple candidate
of such devices is a two-terminal QD thermoelectric device, i.e., a QD
connected with the source (of temperature $T_h$) and the drain (of
temperature $T_c<T_h$) electrodes via resonant tunneling. We adopt
parameters comparable with those in the previous section: the
tunneling rate between the QD and the source (drain) is
$\gamma_e=10$~meV and the QD energy is $E_0=2k_BT_c$. There is
parasitic phonon heat conduction between the source and the drain 
as described by Eq.~(\ref{pr}) and characterized by the two parameters,
$\alpha=0.1$ and $E_{cut}=100$~meV.

We calculate the electrical and heat currents using the Landauer
formula with the energy-dependent transmission function 
\be
{\cal T}_e(E)=\frac{\gamma_e^2}{(E-E_0)^2 + \gamma_e^2} .
\ee
The currents are given by
\begin{subequations}
\begin{align}
& I_e = e\int \frac{dE}{2\pi} {\cal T}_e(E) [f_S(E) - f_D(E)] , \\
& I_{Q,e} = \int \frac{dE}{2\pi} (E-\mu_S) {\cal T}_e(E) [f_S(E) - f_D(E)] , \\
& Q_{in} = I_{Q,e} + I_{Q,pr}, 
\end{align}
\end{subequations}
The energy efficiency and output power can be obtained using the above
currents. 

{As shown in Figs.~3(a) and 3(b), the maximum efficiency for the
elastic thermoelectric device within the linear-approximation
[i.e., Eq.~(\ref{linear})], is 15\% of the Carnot efficiency
and the maximum power is 0.14~nW. Both of them are larger than those
of the inelastic thermoelectric device under the
linear-approximation. However, here the nonlinear transport effect
only leads to marginal improvement of the maximum efficiency and
power: the efficiency is increased only to 17\% of the Carnot
efficiency, while the maximum power is raised only to 0.25~nW. Judging
from the linear-response thermoelectric transport coefficients and the
figure of merit (which is well-defined for the linear-response
regime), the elastic thermoelectric device has much higher efficiency
and power than the inelastic thermoelectric device. However, in the
nonlinear transport regime, the inelastic thermoelectric device has
the same maximum efficiency and larger maximum power. The nonlinear
transport effect leads to much strong enhancement of the maximum
efficiency and power for the inelastic device as compared to the
elastic device.}

{To further explore the nonlinear effects in thermoelectric
transport in the elastic tunneling device, we plot the short-circuit electrical
and heat currents as functions of $T_h/T_c$ in Fig.~3(c). The results
here indicate that the enhancement of the electrical and heat currents
due to the nonlinear transport effect is much weakened for the elastic
device, as compared to the inelastic device. Fig.~3(d) shows that to achieve 20\% deviation between the
``nonlinear'' and the ``linear'' electrical currents, $T_h$ must be
greater than $1.2T_c$. The overall increment of the currents and
voltage due to the nonlinear transport effect is much reduced here, as
compared to the inelastic device.}

{These results are in fact consistent with the findings in the
literature. In Ref.~\cite{robust}, it was found that the
linear-transport approximation is robust for a range of temperature 
biases in resonant tunneling QD thermoelectric devices.}

\begin{figure}[htb]
  \includegraphics[width=7cm]{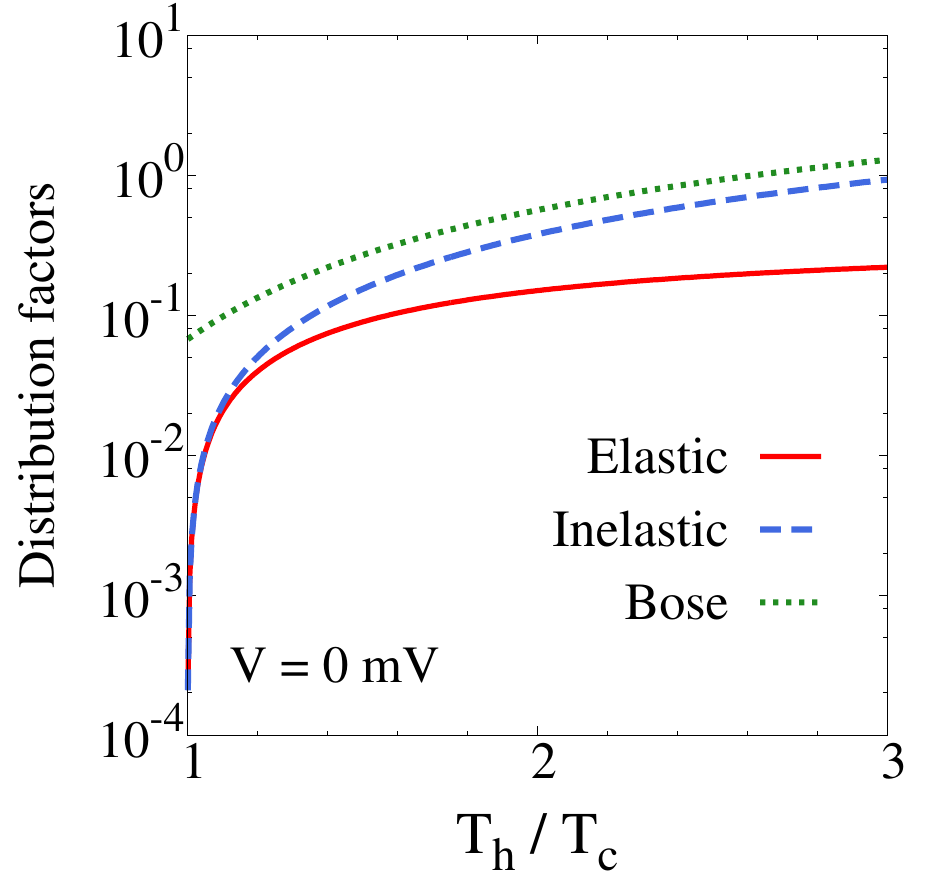}
  \caption{(Color online) Distribution factor that determines the
    magnitude of the charge and heat currents in the
    resonant-tunneling QD device (red solid curve, labeled as
    ``Elastic''), that for the inelastic hopping transport in the
    three-terminal device (blue dashed curve, labeled as
    ``Inelastic''), and the Bose-Einstein distribution factor
    $n_B(|E_2-E_1|/k_BT_h)$ for the inelastic hopping transport in the 
    three-terminal device (green dotted curve, labeled as
    ``Bose''). Parameters: $E_0=2k_BT_c$, $E_1=-E_2=-2k_BT_c$,
    $\mu_S=\mu_D=0$ (i.e., $V=0$). For the two-terminal elastic
    device, $T_S=T_h$ and $T_D=T_c$, whereas for the three-terminal
    inelastic device, $T_S=T_D=T_c$ and $T_P=T_h$. The distribution
    factors are adjusted so that the ``elastic'' and ``inelastic''
    curves start at the same value for very small $T_h/T_c-1$.}
\end{figure}

\section{Distinction between inelastic and elastic thermoelectricity
  from the distribution factor}

To understand the underlying mechanism that leads to the distinction
between the elastic and the inelastic thermoelectric energy
conversions. We study the distribution factors which crucially
determines the magnitude of the charge current and a part of the heat
current. For the two-terminal elastic thermoelectric transport the
charge and heat currents are dominated by the Fermi-Dirac distribution
factor of $f_S(E_0) - f_D(E_0)$ with $T_S=T_h$ and $T_D=T_c$. For the 
three-terminal inelastic thermoelectric transport the currents are
determined by the distribution factor,
$f_1 (1-f_2) N_P^+ - f_2(1-f_1) N_P^-$ with $T_S=T_D=T_c$ and
$T_P=T_h$, which is a mixture of Fermi-Dirac and Bose-Einstein
distributions. When $\gamma_e\gg \Gamma_{12}$\cite{nonlinear}, we can
adopt the approximation $f_1\simeq f_S(E_1)$, $f_2\simeq f_D(E_2)$,
and calculate the distribution factor $f_1 (1-f_2) N_P^+ - f_2(1-f_1)
N_P^-$ easily. We plot the distribution factors for the elastic and
inelastic devices together in Fig.~4. We have adjusted their values so
that they are equal when $T_h/T_c-1$ is very small. Fig.~4 shows
that the distribution factor for the elastic device {\em saturates} at
large temperature bias, whereas the distribution factor for the
inelastic device does not. The saturation of the
distribution factor for the elastic transport is understandable, since
there is a strict bound on the distribution factor $|f_S(E_0) -
f_D(E_0)|\le 1$, due to the Pauli exclusion principle. In contrast,
there is {\em no} bound on the distribution factor for the
inelastic transport. In fact, the Bose-Einstein distribution 
$N_P^+=1/[\exp(\frac{E_2-E_1}{k_BT_h})-1]$ is {\em not} bounded (see the
dotted curve in Fig.~8). This term dominates the rapid growth of the
inelastic distribution factor when the temperature ratio $T_h/T_c$ is
large. This observation explains well the distinction between the nonlinear
effects in the elastic and inelastic devices.

\begin{figure}[htb]
  \includegraphics[width=4.3cm]{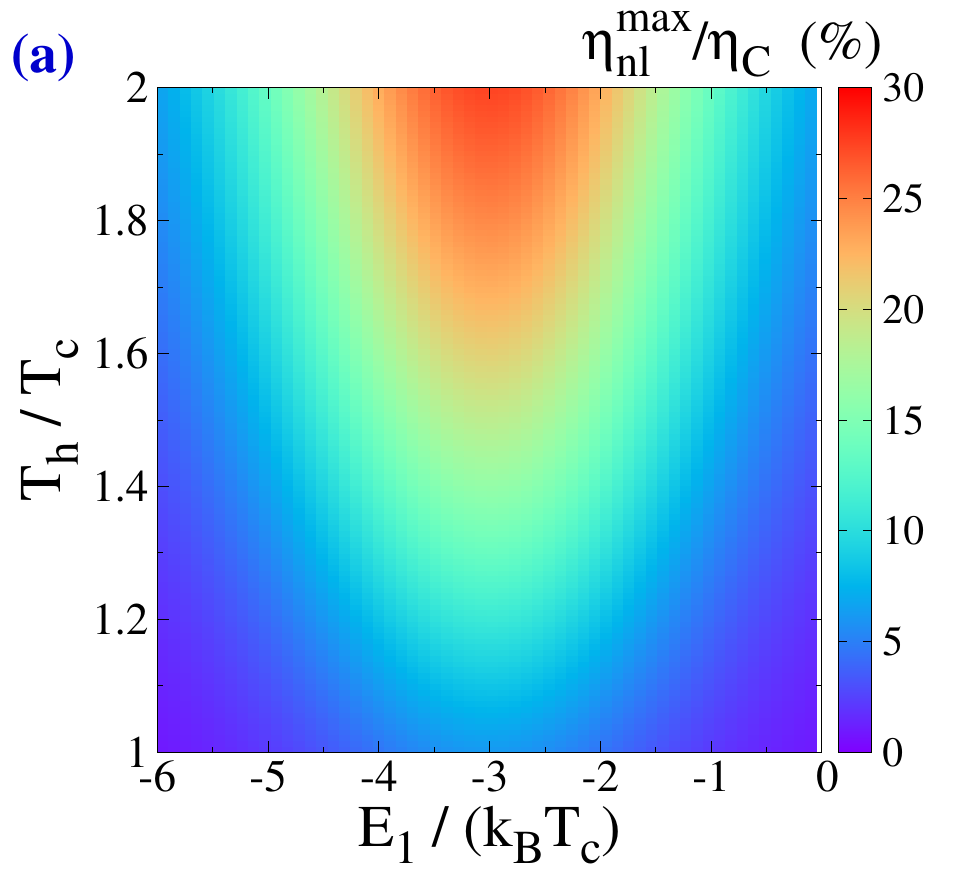}\includegraphics[width=4.3cm]{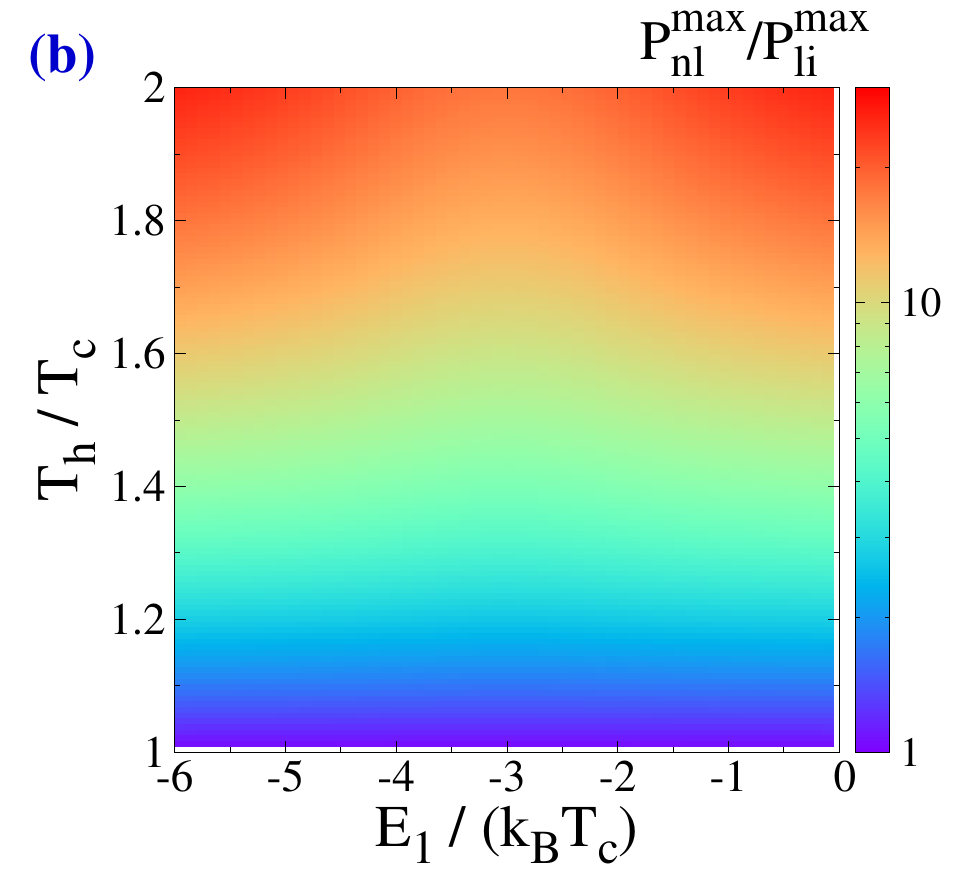}
  \includegraphics[width=4.3cm]{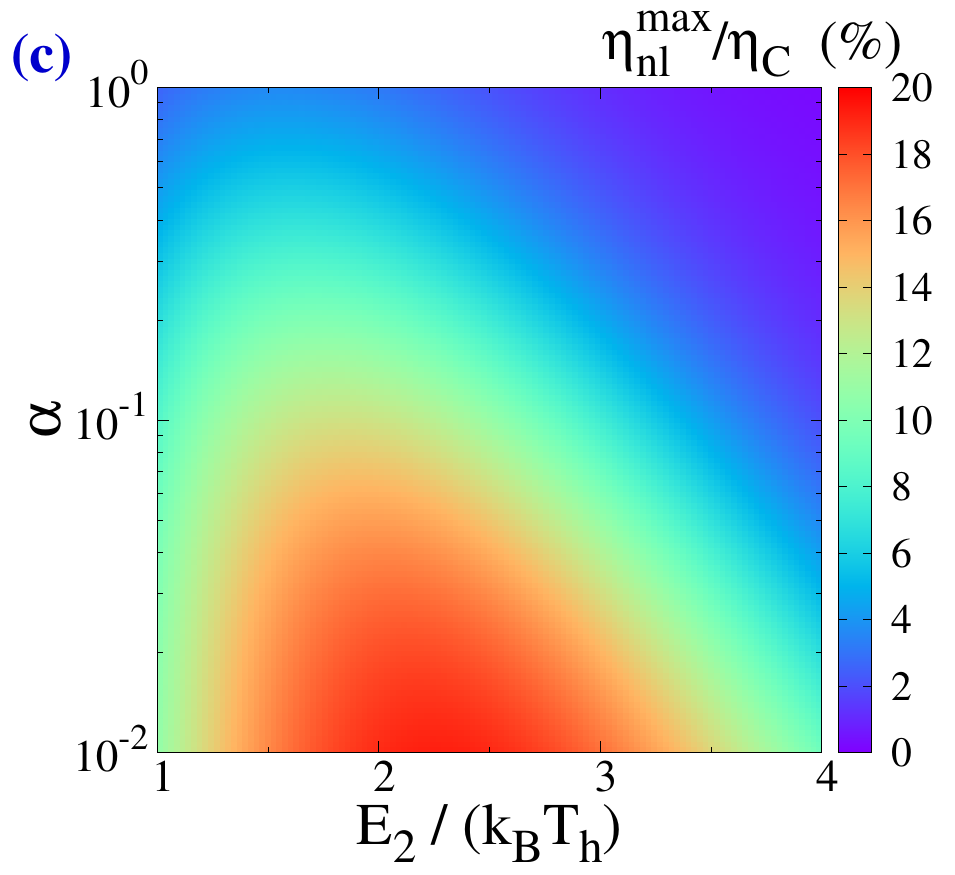}\includegraphics[width=4.3cm]{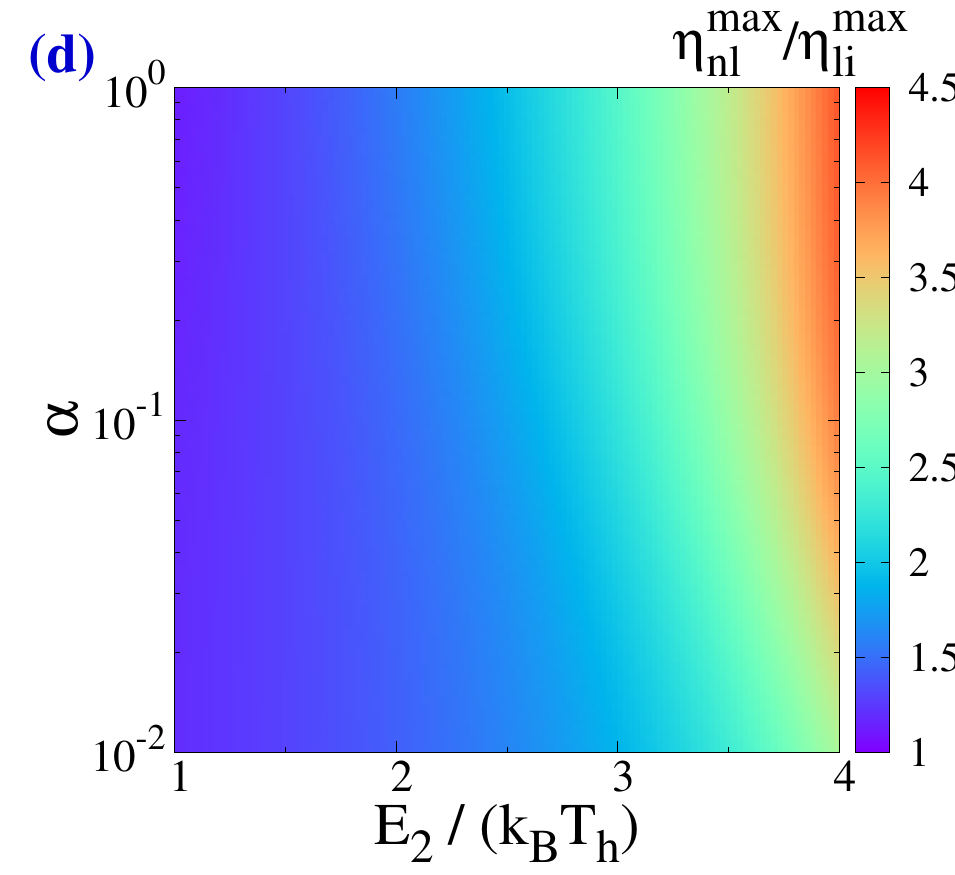}
  \includegraphics[width=4.3cm]{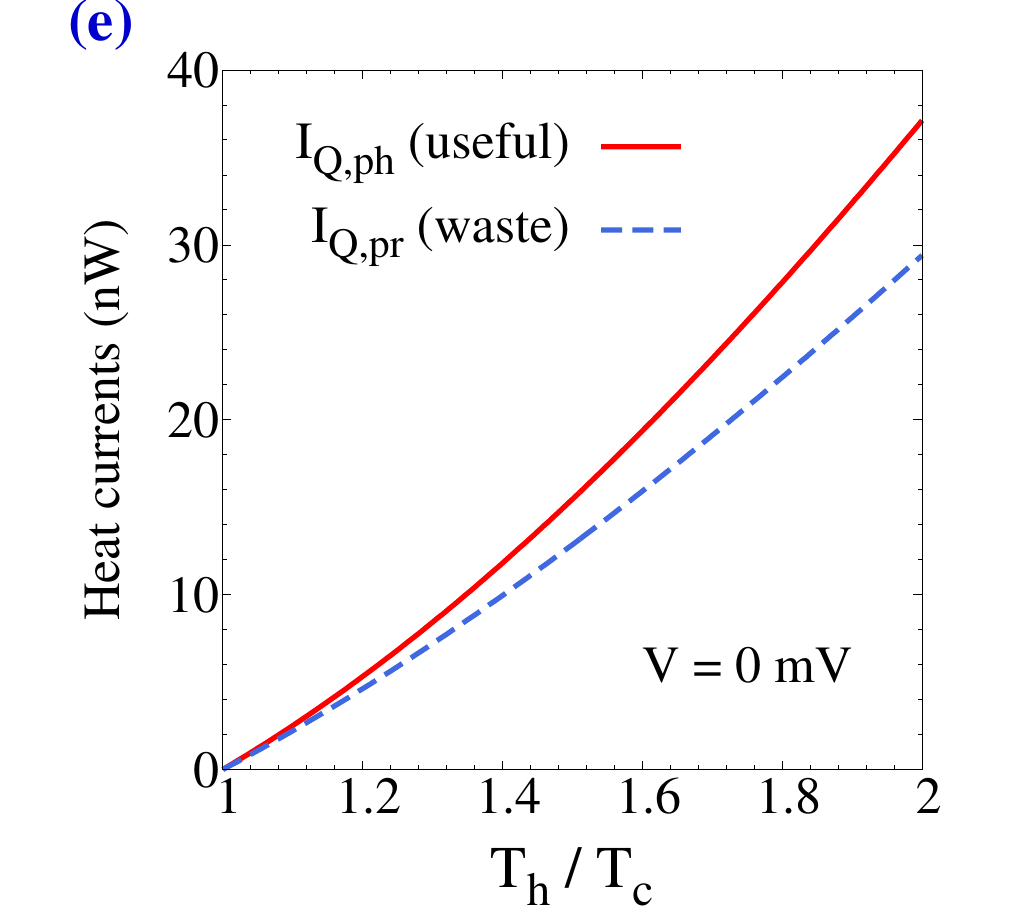}\includegraphics[width=4.3cm]{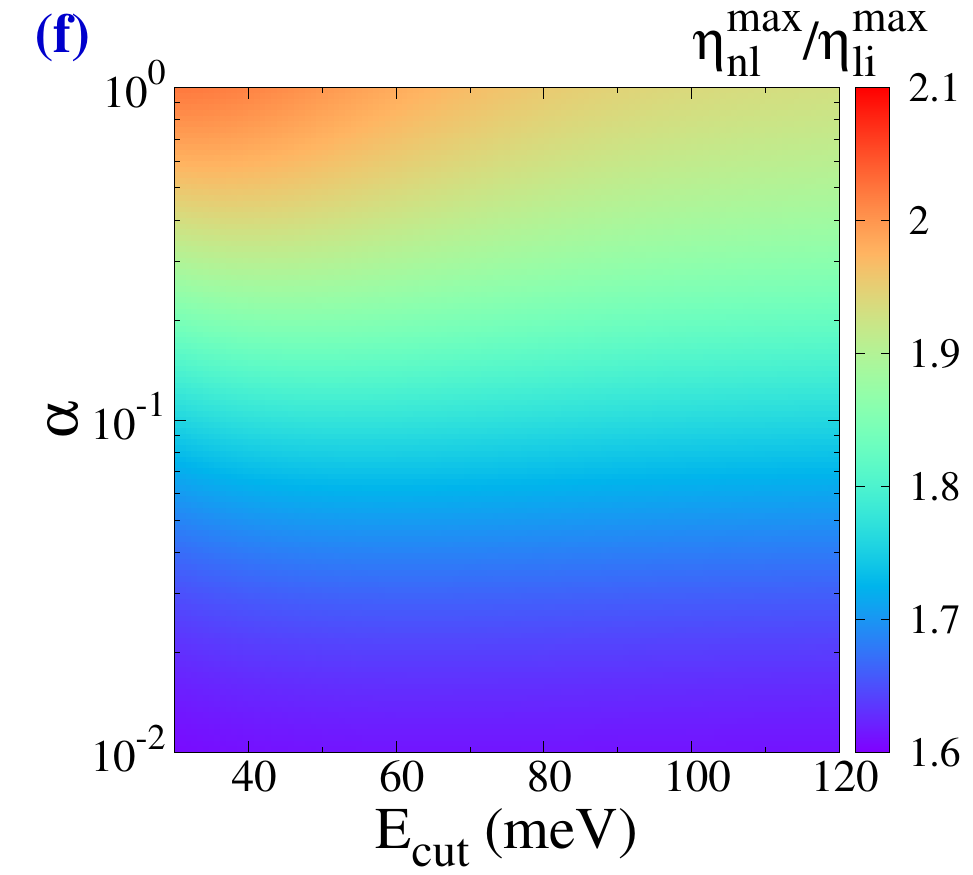}
  \caption{{(Color online) Inelastic thermoelectricity. (a)
    Energy efficiency $\eta_{nl}^{max}/\eta_C$ and (b) power ratio
    $P^{max}_{nl}/P^{max}_{li}$ vs. QD energy $E_1$ (in unit of
    $k_BT_c$) and the temperature of the phonon bath $T_P=T_h$ for
    $E_2=E_1+6k_BT_c$ with $T_c=150$~K. Here the quantities obtained via
    full calculation are denoted by the subscript ``nl'', whereas
    those obtained via linear-approximation are labeled by the
    subscript ``li''. (c) The energy efficiency
    $\eta_{nl}^{max}/\eta_C$ and (d) the ratio
    $\eta_{nl}^{max}/\eta_{li}^{max}$ vs. the QD energy with
    $E_2=-E_1$ (in unit of $k_BT_h$) and the parasitic heat conduction
    (controlled by $\alpha$) 
    for $T_h=1.2T_c=180$~K and $T_c=150$~K. (e) The useful
    thermoelectric heat current $I_{Q,ph}$ and the waste parasitic
    heat current $I_{Q,pr}$ as functions of temperature ratio
    $T_h/T_c$ for $E_2=-E_1=3k_BT_c$ and $V=0$. (f) The ratio $\eta_{nl}^{max}/\eta_{li}^{max}$
    vs. the parasitic heat conduction parameters $\alpha$ and $E_{cut}$ for
    $T_h=1.2T_c=180$~K ($T_c=150$~K) and $E_2=-E_1=3k_BT_c$. The Other
    parameters are $\gamma_e=10$~meV, $\gamma_{ep}=5$~meV, and
    $t=10$~meV. }}
\end{figure}

\section{Variation of temperature bias, quantum-dot energy, and parasitic
  heat conduction}

In the above, by focusing on two specific examples, we have illustrated the
essential difference between inelastic and elastic
thermoelectric devices in the nonlinear transport regime. We revealed
the crucial roles of the Bose and Fermi distributions in the nonlinear
energy efficiency in these two types of thermoelectric devices. To
avoid being misled by particular choice of parameters, in this
section we study the effects of nonlinear transport on thermoelectric
energy efficiency and output power for various temperatures, QD energies, and
parasitic heat conduction.

{We first study how the inelastic thermoelectric efficiency varies with
these parameters. In Fig.~9(a), we show the ratio of the maximum
efficiency over the Carnot efficiency $\eta_{max}/\eta_C$ for various the
temperature $T_h=T_P$ and QD energy $E_1$ when the energy difference
$E_2-E_1$ is fixed to $6k_BT_c$. For each configuration, the energy
efficiency is optimized by varying the voltage $V$. If the
linear-approximation is valid for all $T_h$, then the ratio
$\eta_{max}/\eta_C$ should not vary with the temperature ratio
$T_h/T_c$. The dependence of the ratio $\eta_{max}/\eta_C$ on $T_h/T_c$
directly reflects the nonlinear transport effect. Fig.~5(a) indicates
that the nonlinear effect on the maximum efficiency is prominent: for
$E_1\simeq -3k_BT_c=-E_2$, the maximum efficiency can increase from
6\% to about 27\%! Combining with previous studies on inelastic
thermoelectricity\cite{3tjap,jordan}, we conclude that the ``particle-hole
symmetric'' configuration $E_2=-E_1$ is the optimal configuration for
inelastic thermoelectricity in both linear and nonlinear transport regimes.}

{The enhancement of the maximum power due to the nonlinear
  transport effect, $P^{max}_{nl}/P^{max}_{li}$, is also calculated
  for various QD energies and temperatures [results shown in
  Fig.~5(b)]. Here the enhancement factor,
  $P^{max}_{nl}/P^{max}_{li}$, can be as large as 30. The significant
  enhancement of the maximum power with increasing temperature $T_h$
  is the driving force for the prominent improvement of the maximum
  efficiency shown in Fig.~5(a). Noticeable improvement 
  ($\sim 70\%$) of the maximum power can already be achieved at
  $T_h/T_c$ as small as 1.1. Beside the rapid increase with
  the temperature $T_h$, the enhancement of the maximum power showed no
  considerable dependence on the QD energy.}

{Under the consideration that smaller temperature
  biases are probably more attainable\cite{glushk}, we also study the maximum
  efficiency and its enhancement at $T_h-T_c=0.2T_c$ with
  $T_c=150$~K for the ``particle-hole symmetric'' configuration
  $E_2=-E_1$. The results in Fig.~5(c) indicate that the
  energy efficiency is optimized for small $\alpha$ and $E_2\simeq
  2.5k_BT_h$. This result is slightly different from the previous
  results of $E_2=-E_1\simeq 3k_BT_c$ for the optimization of energy
  efficiency in the linear-response regime\cite{jordan,n1-3t,3tjap}.}

{We further explore the enhancement of the maximum
  efficiency due to the nonlinear transport effect for the
  ``particle-hole symmetric'' configuration $E_2=-E_1$. The
  enhancement of the maximum efficiency is characterized 
  by the ratio $\eta^{max}_{nl}/\eta^{max}_{li}$ in Fig.~5(d)
  (subscripts ``nl'' and ``li'' denote the full calculation with the
  nonlinear effect and the linear-approximation, respectively). The
  enhancement factor $\eta^{max}_{nl}/\eta^{max}_{li}$ can reach close
  to 3 even for a small temperature bias $T_h-T_c=0.2T_c$ with
  $T_c=150$~K. Such enhancement is more prominent for large $E_2$. We
  find that this is because the useful thermoelectric heat current
  surpasses the waste parasitic heat current more significantly when
  the QD energy $E_2=-E_1$ is large. As shown in Fig.~5(e) for one
  typical case,  the useful heat current increases faster than the
  waste heat current with increasing temperature bias, i.e., the
  parasitic heat conduction becomes less and less important with
  increasing temperature bias. Since the enhancement of the
  thermoelectric heat current $I_{Q,ph}=2(E_2-E_1)\Gamma_{12}$
  increases substantially with $E_2$, the
  improvement of the maximum efficiency is more prominent for large
  $E_2$.}

{The results in Fig.~5(f) show that
  the parameter $E_{cut}$ affect the enhancement of energy efficiency
  much weaker than the other parameter $\alpha$. This result is consistent
  with the physics picture that the tunneling heat conduction is
  dominated by the energy scale around $k_BT_h\simeq
  15$~meV. Fig.~5(f) indicates that considerable improvement of
  thermoelectric efficiency can be achieved for a large parameter
  region even for the small temperature bias $T_h-T_c=0.2T_c$.}

\begin{figure}[htb]
  \includegraphics[width=4.3cm]{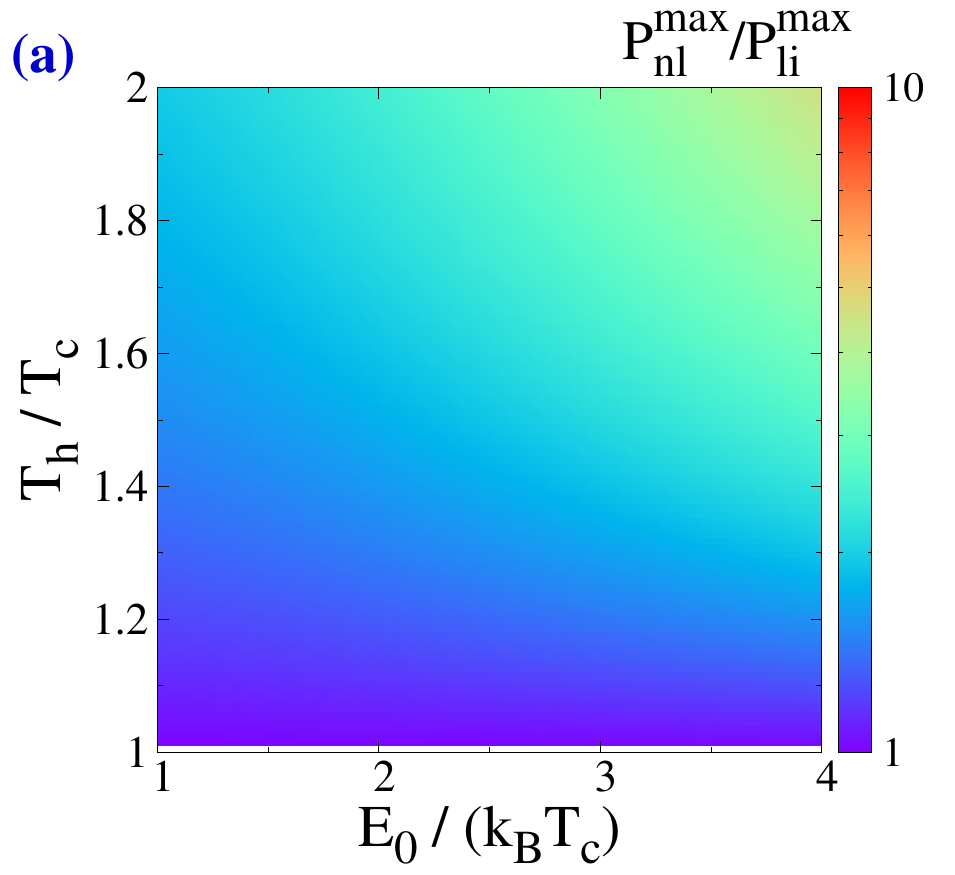}\includegraphics[width=4.3cm]{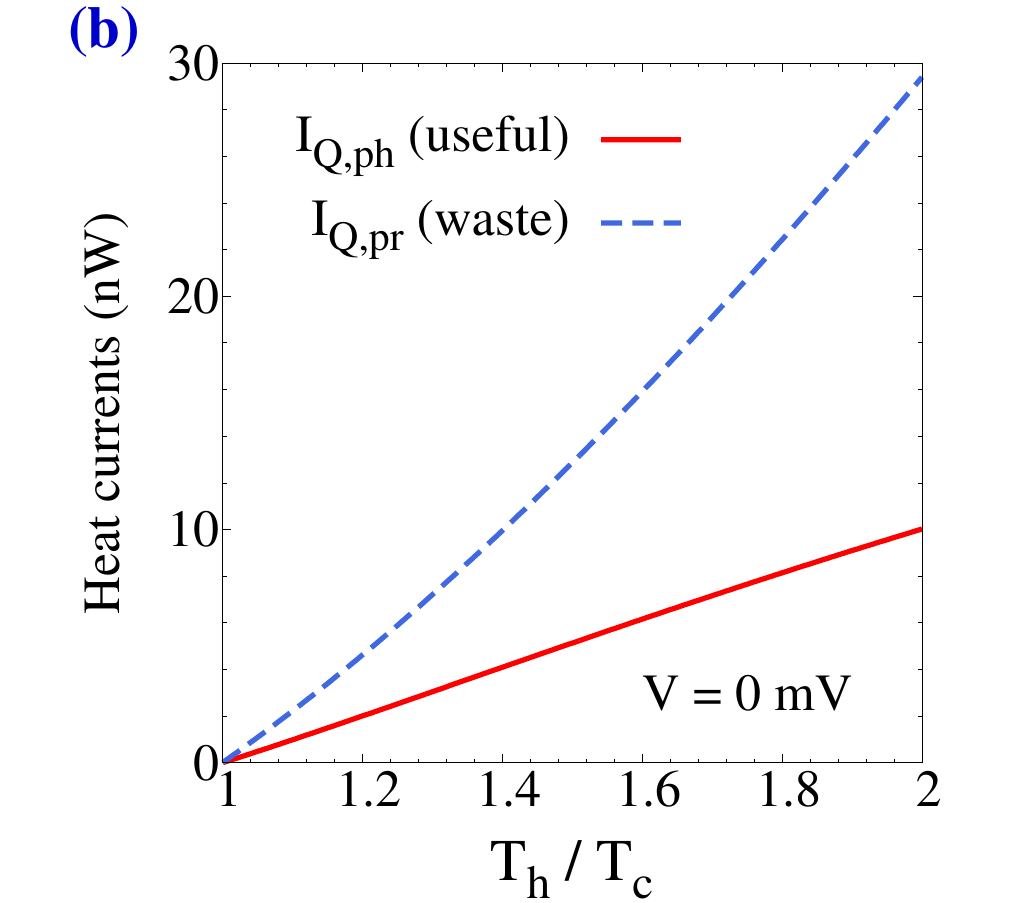}
  \includegraphics[width=4.3cm]{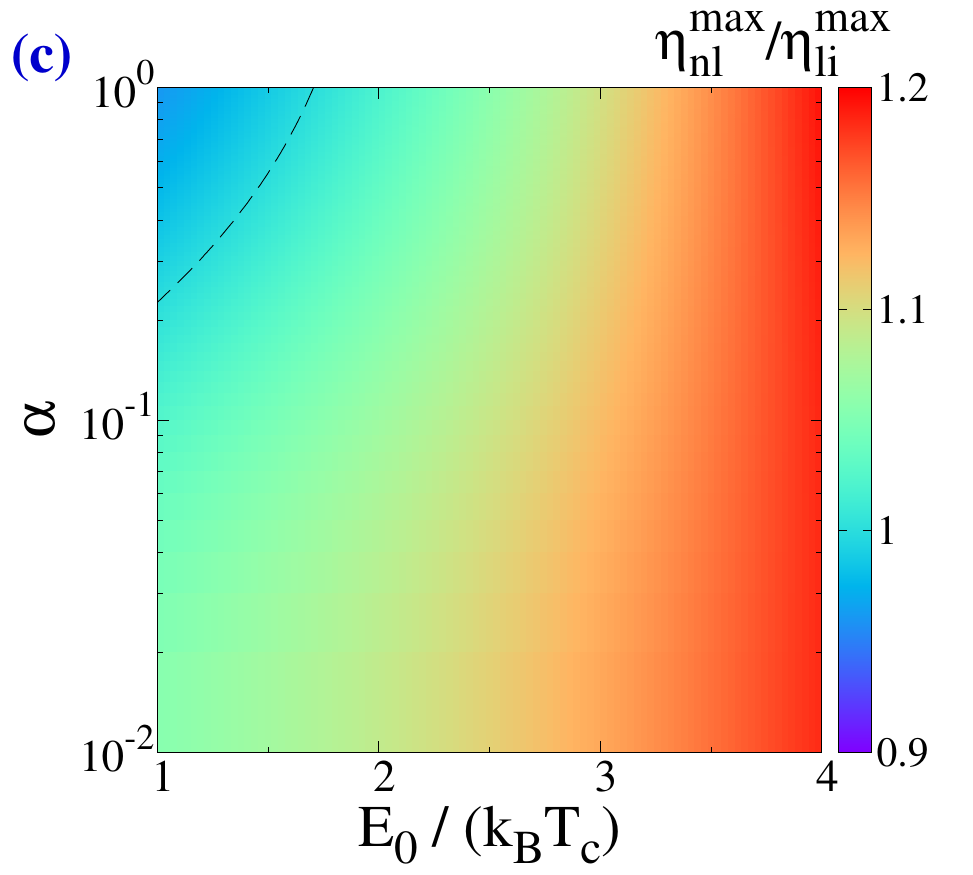}\includegraphics[width=4.3cm]{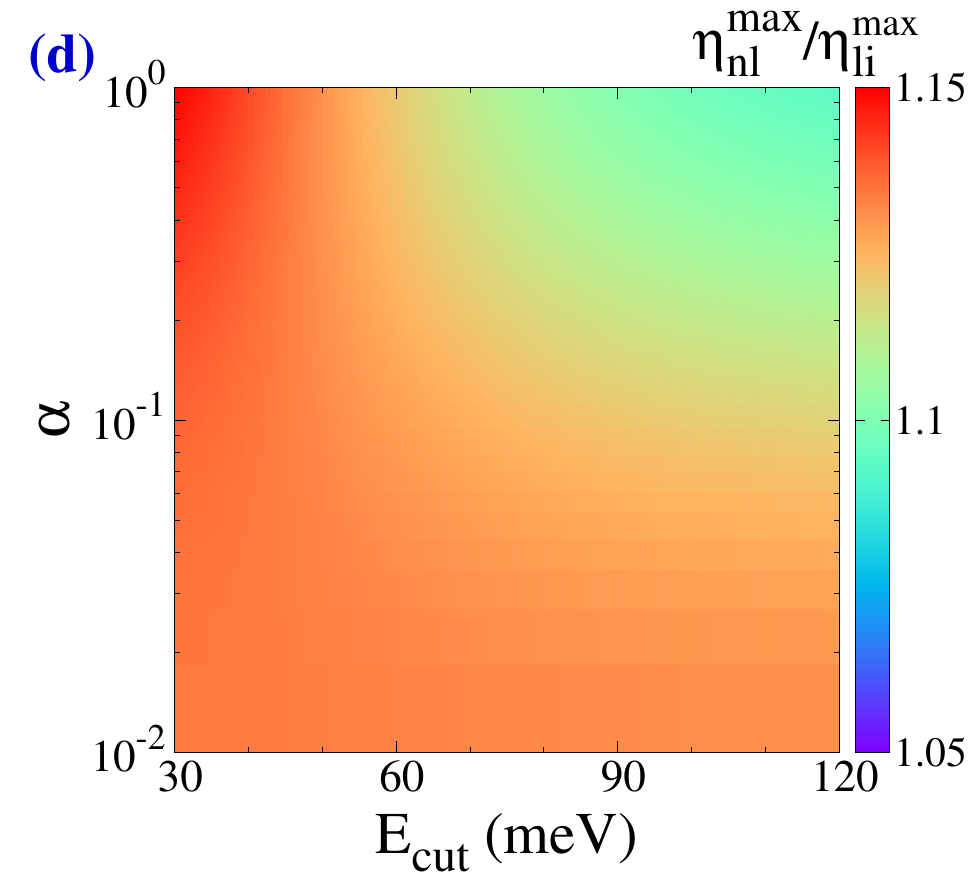}
  \caption{ (Color online) { Elastic thermoelectricity. (a) The ratio of
    the maximum powers $P^{max}_{nl}/P^{max}_{li}$ as a function of
    the QD energy $E_0$ (in unit of $k_BT_c$) and the temperature
    ratio $T_h/T_c$ for $T_c=150$~K. Here the quantities obtained via the full
    calculation (including the nonlinear effect) are labeled with the
    subscript ``nl'', whereas those obtained via the
    linear-approximation is denoted with the subscript ``li''. (b) The
    useful thermoelectric heat current $I_{Q,e}$ and the waste
    parasitic heat current $I_{Q,pr}$ vs. the temperature ratio
    $T_h/T_c$ for $E_0=3k_BT_c$, $T_c=150$~K, and $V=0$. (c) The ratio
    $\eta_{nl}^{max}/\eta_{li}^{max}$ as a function of the QD energy
    $E_0$ and the parasitic heat conduction parameter $\alpha$ for 
    $T_h=1.2T_c=180$~K ($T_c=150$~K) and $E_{cut}=100$~meV. The
    black-dashed curve represents the contour of
    $\eta_{nl}^{max}/\eta_{li}^{max}=1$. (d) The ratio
    $\eta_{nl}^{max}/\eta_{li}^{max}$ vs. the parasitic heat
    conduction parameters $\alpha$ and $E_{cut}$ for 
    $T_h=1.2T_c=180$~K ($T_c=150$~K) and $E_0=3k_BT_c$. For all
    figures here $\gamma_e=10$~meV, $\alpha=0.1$ and
    $E_{cut}=100$~meV, unless specified.}}
\end{figure}

{We now turn to the elastic tunneling QD thermoelectric
devices. We start with the enhancement of the maximum power due to the
nonlinear transport effect. As shown in Fig.~6(a), the increase of the
maximum power due to the nonlinear effect is much weakened. The ratio
$P^{max}_{nl}/P^{max}_{li}$ here is smaller than 5, while the ratio
for the inelastic thermoelectric device can be as large as $\simeq 30$
[see Fig.~5(b)]. Besides, Fig.~6(b) shows that the waste parasitic
heat current increases faster with the temperature $T_h$ than the
useful thermoelectric heat current. This result indicates that the
parasitic heat conduction becomes more and more important in the
nonlinear transport regime for elastic thermoelectric devices.}

{Fig.~6(c) gives the efficiency enhancement factor
$\eta^{max}_{nl}/\eta^{max}_{li}$ for various QD energy $E_0$ and
parasitic heat conduction (as controlled by $\alpha$) at the same
temperature bias of $T_h-T_c=0.2T_c$ as in Fig.~5(d). It is seen that
for a large parameter region, the improvement of the maximum
efficiency is quite small. Besides, there are regimes where the
nonlinear effect reduces the maximum efficiency. The overall
improvement of the maximum efficiency due to the nonlinear transport
effect is much weaker in elastic thermoelectricity, as compared to
inelastic thermoelectricity. This conclusion is also supported by
Fig.~6(d) where the enhancement factor
$\eta^{max}_{nl}/\eta^{max}_{li}$ only reaches to $\le 1.16$, i.e.,
improvement of the maximum efficiency only by $\le 16\%$. This value
is much smaller than the improvement of $\le 110\%$ for the efficiency
of the inelastic thermoelectric devices [see Fig.~5(f)]. The results
in Figs.~6(c) and 6(d) are in consistent with Figs.~6(a) and 6(b): the
small improvement in the maximum power is compensated by the strong
increase of the parasitic heat conduction, leading to marginal
improvement or even reduction of the maximum efficiency.}

\section{conclusions and discussions}
{We have performed a comparative study of the nonlinear transport
effect on the maximum efficiency and power for inelastic and elastic
thermoelectric devices systematically. We find that the nonlinear
effect can significantly improve the performance of inelastic
thermoelectric devices, whereas it only has marginal
effects on the performance of elastic tunneling thermoelectric
devices. The latter observation is, in fact, consistent with previous
studies\cite{robust}. The former, however, is found only in this
work.}

{We revealed that the underlying mechanism of such distinction is due to
the fact that the distribution factor, particularly the Bose distribution,
involved in the inelastic transport is much more prominently increased
under considerable temperature or voltage bias. Unlike Fermi
distributions, the Bose distribution is not bounded, which leads to
significant enhancement of the inelastic thermoelectric currents and
voltages, and then the maximum efficiency and power. It is also
interesting to notice that, in the literature, it was also proposed
that small temperature bias is benificial for efficiency for a
particular type of thermoelectric device\cite{olsen}, which is
opposite to our findings here for inelastic thermoelectricity.}

{We remark that the numbers (temperatures, energy, and tunneling rates)
in this study can be modified when adopt to certain circumstances. Our
study is not restricted to specific materials, where our conclusions
may still hold true beyond such materials, since the underlying
mechanism is quite robust. We confirm our
conclusion with a systematic survey of various control parameters as
well. Nevertheless, strong electron-phonon interaction is needed to
form powerful and efficient inelastic thermoelectric devices. We
propose to use ionic crystals such as GaN where the optical phonon
frequency and Debye temperature is close to $90$~meV, and the
electron-phonon interaction is very strong (even much stronger than in
GaAs)\cite{gan,gan2}. Though it is physically speculated that strong
temperature/voltage bias are easier for nanoscale thermoelectric
devices\cite{rev,cooling}, experimental endeavors are still demanded to
achieve such a regime for nonlinear thermoelectricity and the
challenges related to large temperature/voltage
gradients and heat/charge fluxes\cite{cam,exp1,exp2,exp3,prapp}.}

We must emphasize that to illustrate the principle, we treated here a
single pair of QDs, where their level separation, $\ome=E_2-E_1$, is
given. The efficiency is defined with the well-defined energy $\ome$
absorbed from the boson source into the electron system. Obviously, in
such an analog of the solar cell only a small part of the input energy
flux (of phonons) is used. This can be remedied by considering an
ensemble of self-assembled QDs\cite{3tjap}, or using the continuous
spectrum of, say, a $p$-$n$ junction with very small band gap, in the
manner of Ref.~\cite{pn}.

{In the regime of very large temperature bias, such as for solar cells,
$T_h\simeq 5800$~K and $T_c\simeq 300$~K, the enhancement of nonlinear
transport can become very significant. The maximum power and
efficiency also benefits from the continuous spectrum and large
density of states for optical absorption, as well as the frequency
filtering by the band gap [Eq.~(1) indicates that the efficiency is
greater if the variance of the frequencies of the absorbed photons are
smaller]. The solar cell efficiency is optimized around a band gap of
1.3~eV\cite{solar,shockley,booksc}, i.e., about $2.6k_BT_h$, which is in good agreement with our
numerical estimation [see Fig.~5(c)], though our calculation is based
on a simple inelastic transport model based on a pair of QDs. Thus our
study provides a qualitative understanding on why the energy
efficiency of solar cell is much better than conventional thermoelectric
devices based on elastic transport. In the analog between inelastic
thermoelectricity and solar cell, efficient and powerful
thermoelectric device can be achieved via an ``energy gap''
$\ome\simeq 80$~meV for $T_h=400$~K with strong electron-phonon
interaction, which may be achieved using GaN QDs.} The similarities
between solar cells and inelastic thermoelectric devices  
suggest the promising future of inelastic thermoelectricity.

\section*{Acknowledgment}
JHJ acknowledges supports from the National Science
Foundation of China (no. 11675116) and the Soochow university faculty start-up
funding. He also thanks Weizmann Institute
of Science for hospitality and support. YI acknowledges support from
the Israeli Science Foundation (ISF), the US-Israel Binational Science
Foundation (BSF), and the Weizmann Institute.

{}

\end{document}